\documentclass[aps,twocolumn,prl,showpacs]{revtex4}

\usepackage{graphicx}

\begin{document}

\title{Uniformly frustrated $XY$ model without vortex-pattern ordering}
\author{S. E. Korshunov}
\affiliation{L. D. Landau Institute for Theoretical Physics,
Kosygina 2, Moscow 119334, Russia}
\date{September 22, 2004}

\begin{abstract}
The uniformly frustrated $XY$ model with $f=1/3$ on a dice lattice
is shown to possess a so well developed accidental degeneracy of
its ground states that the difference between the free energies of
fluctuations does not lead to the stabilization of a particular
vortex pattern down to zero temperature. Nonetheless, at low
temperatures the system is characterized by a finite helicity
modulus whose vanishing (at a finite temperature) is related with
the dissociation of half-vortex pairs.
\end{abstract}

\pacs{74.81.Fa, 64.60.Cn, 05.20.-y}

\maketitle

It is well known that in the presence of external magnetic field a
regular two-dimensional Josephson junction array can be described
by the Hamiltonian of a uniformly frustrated $XY$ model
\cite{TJ-L}:
\begin{equation}                                \label{H}
H=-J\sum_{(\bf jk)}\cos(\varphi_{\bf k}-\varphi_{\bf j}-A_{\bf
jk})\;,
\end{equation}
where $J$ is the Josephson coupling constant, the fluctuating
variables $\varphi_{\bf j}$ are the phases of the order parameter
on superconducting grains forming the array, the quenched
variables $A_{\bf jk}$ are determined by the vector potential of
the field and the summation is taken over all pairs of grains
connected by a junction. The form of Eq. (\ref{H}) assumes that
the currents in the array are sufficiently small, so their proper
magnetic fields can be neglected.

The directed sum of the variables $A_{\bf jk}\equiv -A_{\bf kj}$
along the perimeter of each plaquette should be equal to $f$, the
ratio of the magnetic flux per plaquette to the flux quantum
$\Phi_0$.  It is sufficient to consider the interval
\makebox{$f\in[0,\frac{1}{2}]$}, because all other values of $f$
can be reduced to this interval by a simple replacement of
variables \cite{TJ-L}. The case of $f=0$ corresponds to the
absence of frustration.

For rational $f$  the nature of the vortex pattern in the
low-temperature phase of a uniformly frustrated $XY$ model is
usually unambiguously determined by the structure of its ground
states. The best known examples belonging to this class are the
models with $f=1/2$ and square or triangular lattice \cite{ffx}.
On the other hand, quite often the ground states of a uniformly
frustrated $XY$ model are characterized by an accidental
degeneracy not related to symmetry \cite{K86,KVB}. In such cases
the nature of vortex ordering at low temperatures cannot be
determined without comparing the free energies of fluctuations in
the vicinities of different degenerate ground states \cite{VBCC}.

Quite often (for example, when the lattice is triangular and $f
=1/3$ or $1/4$ \cite{KVB}), this mechanism of the removal of an
accidental degeneracy  works already at the harmonic level, but in
some situations (at $f=1/2$ on honeycomb, dice and {\em kagome}
lattices) one has to take into account the anharmonicites
\cite{fxh}. However, in all the cases investigated insofar, at low
enough temperatures a particular periodic vortex pattern can be
expected to be stabilized by fluctuations (in the thermodynamic
limit).

In the present work we demonstrate that the uniformly frustrated
$XY$ model with $f=1/3$ and dice lattice (see Fig. 1) has rather
unique properties. Namely, this is the first example of a
frustrated $XY$ model in which the accidental degeneracy of ground
states is so well developed that vortex pattern remains disordered
at arbitrarily low temperature, although the free energies of
fluctuations are different for different periodic patterns. As a
consequence, the only phase transition which takes place in this
system with the increase of temperature is related to dissociation
of pairs of fractional vortices with topological charges $\pm
1/2$. A possibility for the  stabilization of a specific periodic
pattern can appear only if one goes beyond the limits of the $XY$
model. In the conclusion we briefly discuss the removal of the
accidental degeneracy by magnetic interactions and its possible
consequences.

The interest to magnetically frustrated superconducting systems
with a dice lattice have been motivated by the unusual properties
of a single electron spectrum in this geometry \cite{VMD}. In
recent years superconducting wire networks and Josephson junction
arrays with a dice lattice have become the subject of active
experimental investigations \cite{Abil,Serr,TTM} and numerical
simulations \cite{CF}. Magnetically frustrated Josephson junction
arrays formed by rhombic plaquettes have been also discussed in
the context of creation of topologically protected qubits
\cite{IF}.

\begin{figure}[b]
\includegraphics[width=55mm]{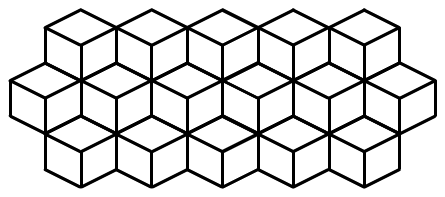}
\caption[Fig. 1] {Dice lattice is periodic and consists of
identical rhombic plaquettes with three different orientations.
}\label{fig1}
\end{figure}

\vspace*{2mm} Since both $\varphi_{\bf j}$ and $A_{\bf jk}$ depend
on a particular choice of the gauge, it is more convenient to
describe different states of the system in terms of the
gauge-invariant phase differences
\[
\theta_{\bf jk}=\varphi_{\bf k}-\varphi_{\bf j}-A_{\bf jk}\equiv
-\theta_{\bf kj}\;,
\]
which below are always assumed to be reduced to the interval
$(-\pi,\pi)$. The subscripts $\bf j$ and $\bf k$ will be used to
denote six- and threefold coordinated sites respectively. If one
defines the variable $m_{\bf jj'}$  to be given by the directed
sum of variables $\theta$ over the perimeter of the plaquette
$\langle\bf jkj'k'\rangle$ divided by $2\pi$, the plaquettes for
which $m_{\bf jj'}$ is equal to $1\!-\!f$ (rather then to $-f$)
are usually referred to as containing vortices. Different local
minima of (\ref{H}) can be then classified by specifying the
positions of vortices, whose concentration for $f=1/3$ should be
exactly equal to one third.

\begin{figure}[b]
\includegraphics[width=80mm]{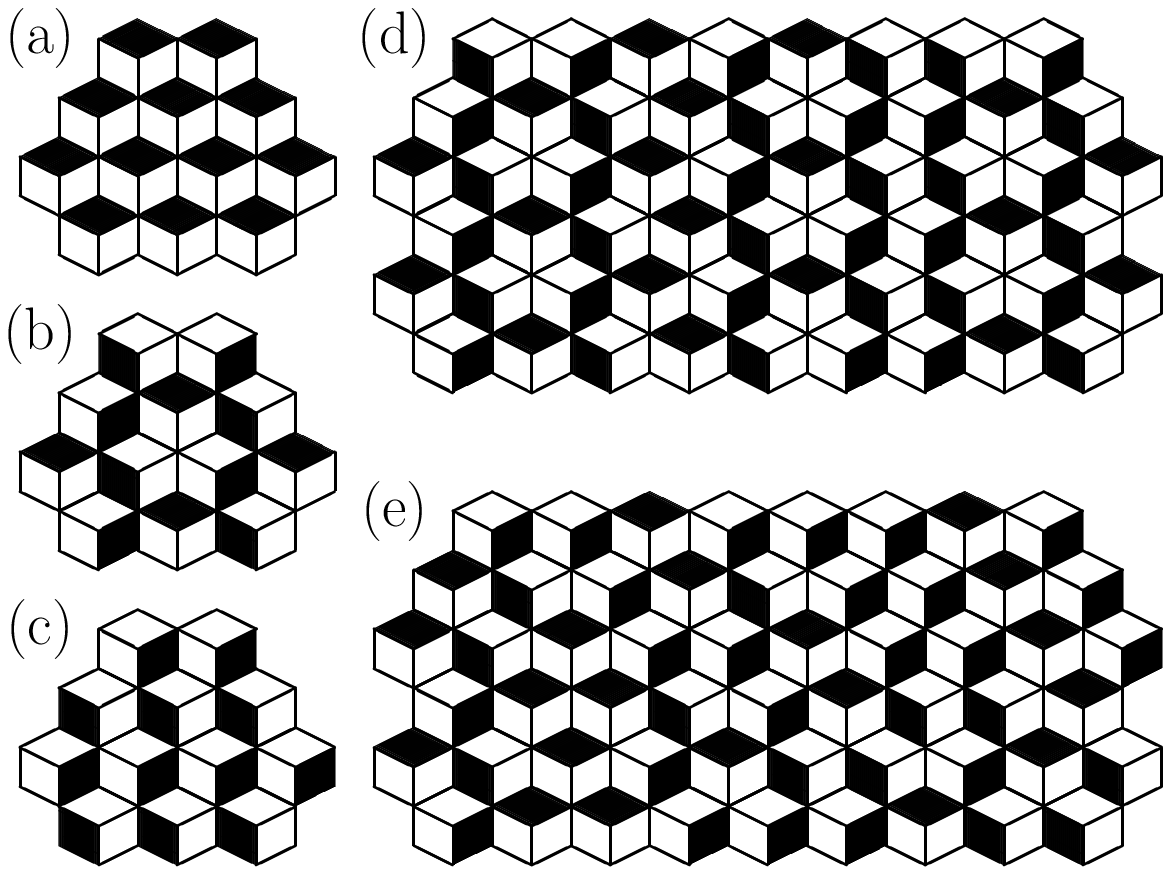}
\caption[Fig. 2] {Some ground states of the frustrated $XY$ model
with a dice lattice and $f=1/3$. The black plaquettes correspond
to $m=2/3$ and white plaquettes to $m=-1/3$.} \label{fig2}
\end{figure}

The remarkable property of the considered $XY$ model is that all
minima of (\ref{H}) corresponding to different vortex
configurations in which vortices do not occupy adjacent plaquettes
have the same energy (per bond), \makebox{$E_0= -(2/3)J$}.  Some
of these states are schematically shown in Fig. 2. In all such
states the variables $\theta$ are equal to $\pi/3$ on all bonds
surrounding a vortex and to zero on all other bonds (shared by two
plaquettes without vortices). It is evident that these states
correspond to the absolute minimum of energy, because they
minimize the energy separately for each plaquette containing a
vortex and for each of the remaining bonds. The barriers between
the ``neighboring"  ground states have the height \makebox{$E_{\rm
b}=6(2-\sqrt{3})J\approx 1.61J$}.

All these ground states [whose degeneracy survives even if the
interaction in (\ref{H}) deviates from cosine] can be put into
correspondence with the ground states of the antiferromagnetic
Ising model defined on the triangular lattice $\cal T$ formed by
the sixfold coordinated sites $\{\bf j\}$. In all ground states of
such a model each triangular plaquette should contain exactly one
bond with parallel spins ($s_{\bf j}s_{\bf j'}=1$) \cite{Wann}.
The existence of a mapping becomes evident as soon as one notices
that in all ground states of the considered $XY$ model each
hexagon formed by three neighboring plaquettes should contain
exactly one vortex. The plaquettes with vortices can be then
identified with the bonds connecting parallel spins by setting
$s_{\bf j}s_{\bf j'}=2m_{\bf jj'}-1/3$.

At zero temperature, $T=0$, the antiferromagnetic Ising model on a
triangular lattice is characterized by an algebraic decay of
correlation functions \cite{Steph} and a finite extensive entropy,
which is related to the possibility of creation of zero-energy
domain walls forming closed loops \cite{Wann}. The same partition
function can be interpreted as the partition function of the SOS
(solid-on-solid) model suitable for the description of the height
fluctuations on the $(111)$ facet of a crystal with a simple cubic
lattice \cite{BH}. In terms of this SOS model zero-energy domain
walls correspond to zero-energy steps, and the algebraic
correlations of spins are translated into logarithmic correlations
of integer variables $h_{\bf j}$ and $n_{\bf k}$, which can be
associated with the height of the surface. These variables can be
introduced following the relations
\[
h_{{\bf j\pm e}_\alpha}=h_{\bf j}\pm 3m_{{\bf j,j\pm e}_\alpha},
~~n_{\bf k}=\frac{1}{3}\sum_{\bf j=j(k)}h_{\bf j}\;,
\]
where ${\bf e}_\alpha$ (with $\alpha=1,2,3$) are the three basic
vectors (${\bf e}_1+{\bf e}_2+{\bf e}_3=0$) of $\cal T$ and $\bf
j(k)$ are the three nearest neighbors of $\bf k$ on the dice
lattice. According to Ref. \onlinecite{BH}, for \makebox{$|{\bf
k}_1-{\bf k}_2|\gg 1$}
\begin{equation}                          \label{corfunc}
\langle(n_{\bf k_1}-n_{\bf k_2})^2\rangle\propto \frac{9}{\pi^2}
\ln |{\bf k_1-k_2}|\;.
\end{equation}

The form of Eq. (\ref{corfunc}) demonstrates that the SOS model is
in the rough phase and that at $T=0$ the large-scales fluctuations
of $n$ can be described by a continuous Gaussian Hamiltonian,
\begin{equation}                            \label{Hh}
H=\frac{K}{2}\int d^2{\bf r}(\nabla n)^2\;,
\end{equation}
where the dimensionless effective rigidity $K$ (which is of
entropic origin) is equal to $K_0=\pi/9$ \cite{BH}. The phase
transition of the SOS model to the smooth phase
% (the roughening transition \cite{Weeks})
would take place when $K=\pi/2$ \cite{NHB}, thus the
system is situated relatively far from the transition point.

The simplest periodic ground state shown in Fig. 2(a) (the striped
state) has been discussed in Ref. \onlinecite{CF}. In terms of the
SOS representation this state has the maximal possible slope,
whereas a flat state of the SOS model (with $n_{\bf
k}=\mbox{const}$) corresponds to the honeycomb vortex pattern of
Fig. 2(b). Fig. 2(d) shows two parallel steps (of opposite signs)
which separate flat states with $\Delta n=\pm 1$. If the left step
of Fig. 2(d) is repeated as often as possible, one obtains the
striped state of Fig. 2(a). On the other hand, the repetition of
the right step of Fig. 2(d) leads to the zigzag state shown in
Fig. 2(c). However, the steps do not have to be straight, and a
typical ground state looks rather disordered, Fig. 2(e).

If the SOS model would be in a smooth phase, the vortex
configuration would be characterized  by the long-range order
corresponding to the formation of the honeycomb pattern of Fig.
2(b). The fluctuations of $n_{\bf k}$ and $h_{\rm j}$ lead to the
replacement of the true long-range order by an algebraic decay of
correlations of $m_{\bf jj'}$,
\[
\langle m_{\bf j_1j'_1}m_{\bf j_2j'_2}\rangle\propto |{\bf
j_1-j_2}|^{-\eta}\;,
\]
where $\eta=2$, as follows from the results of Ref.
\onlinecite{Steph}. However, one can expect this correlation
function to be modulated according to the honeycomb pattern.

At finite temperature, $T>0$, the equivalence between the
considered $XY$ model and the Ising model is no longer exact,
because the $XY$ model (i) allows for the existence of continuous
fluctuations (spin waves) and (ii) has a more complex
classification of topological defects. Numerical calculation of
the integrals over the Brillouin zone which determine the free
energies of harmonic fluctuations in the vicinity of the three
periodic ground states shown in Figs. 2(a), 2(b) and 2(c) reveals
that this free energy is the lowest in the honeycomb state
\cite{footnote1}. The difference between the free energies of
fluctuations in the zigzag and honeycomb states (normalized per
single site of $\cal T$) is given by $\gamma T$, where
$\gamma\approx 2. 27\cdot 10^{-3}$.

Since in terms of the SOS model the zigzag state can be considered
as the sequence of steps with the unit density, the same quantity
can be also used as an estimate for the effective energy of a step
per elementary segment, \makebox{$E_{\rm st}\approx \gamma T$}.
The positiveness of $E_{\rm st}$ should lead to a decrease of
fluctuations of $n_{\bf k}$, but since one always has
\makebox{$E_{\rm st}/T\ll 1$}, this decrease has to be relatively
small. In terms of (\ref{Hh}) the influence of a small step
energy, $E_{\rm st}\approx \gamma T$, is translated into a very
small ($<1\%$) correction to $K_0$, $K=K_0+(2/\sqrt{3})\gamma$.
This definitely leaves the system far from the transition to the
smooth phase, and, therefore, cannot lead to any qualitative
changes from the zero-temperature behavior.

\begin{figure}[b]
\includegraphics[width=48mm]{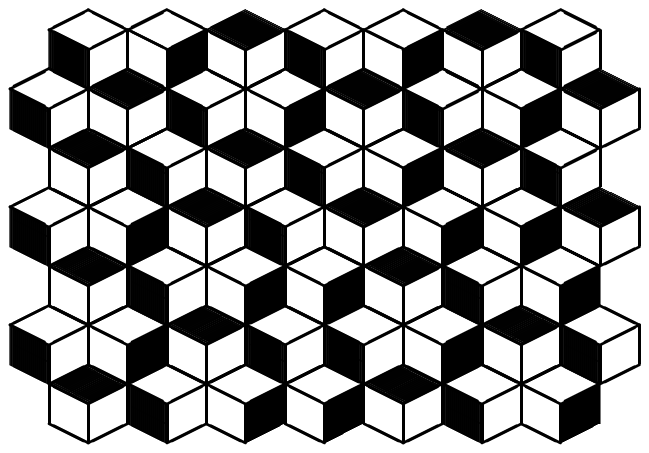}
\caption[Fig. 3] {A possible structure of a half-vortex.}
\label{fig3}
\end{figure}

In terms of the original $XY$ model each step of the SOS model
corresponds to a line whose crossing shifts the phase variables
$\varphi_{\bf j}$  by $\pi$ (with respect to what they would be in
the absence of this step) on one of the three sublattice into
which the triangular lattice $\cal T$ can be split. After crossing
three steps of the same sign the variables $\varphi_{\bf j}$ are
shifted by $\pi$ for all three sublattices. In other words, the
state which is obtained in such a way differs from the original
state by a global rotation of all phases by $\pi$. As a
consequence,  the elementary topological excitation of the
considered $XY$ model is an object where three steps of the same
sign merge together (see Fig. 3) that looks like a screw
dislocation with Burgers number $b=\pm 3$, on going around which
the phase experiences a continuous rotation by $\pi$. For brevity
we shall call such defects half-vortices. Since the phase shift by
$\pi$ can be achieved by a phase rotation in both directions, the
signs of the two topological charges of a half-vortex (the Burgers
number and the vorticity) are not related to each other and can be
arbitrary.

The core of a half-vortex can be associated with the
three-plaquette cluster which instead of containing exactly one
vortex contains either two vortices or no vortices at all (see
Fig. 3). In the framework of the antiferromagnetic Ising model the
analogous defect cannot exist, because each plaquette can contain
only an odd number ($1$ or $3$) of bonds with $s_{\bf j}s_{\bf
j'}=1$. The core energy of a half-vortex should be of the order of
$J$.

In accordance with the double nature of half-vortices their
interaction consists of two contributions of different origin.
Both of them are logarithmic. The first one (the direct
interaction) is related to the energy which is required to create
the phase twist around the cores and is completely analogous to
the interaction of ordinary vortices (with integer vorticity).
This interaction is characterized by the prelogarithmic factor
$P_{\rm V}=(\pi/2)\Gamma$, where $\Gamma$ is the helicity modulus
of the system [at $T=0$ in the honeycomb state
$\Gamma=\Gamma_0\equiv(5/4\sqrt{3})J$]. The second contribution is
of entropic origin and is related to the interaction of
half-vortices as dislocations. It follows from Eq. (\ref{Hh}) that
for this interaction the prelogarithmic factor is given by
\begin{equation}                            \label{PD}
P_{\rm D}(b)=\frac{Kb^2}{2\pi}T\;,
\end{equation}
where one should put $b=3$.

At low temperatures ($T\ll J$) the direct interaction of
half-vortices is dominant, which binds them into small pairs with
zero total vorticity. However, the Burgers number of such a pair
does not have to be zero, but can be also equal to $\pm 6$. This
returns one to the situation in the antiferromagnetic Ising model
in which the elementary topological excitations are the
dislocations with $b=\pm 6$ \cite{NHB,L83}, the only difference
being a slightly larger value of $K$. Substitution of $K\approx
K_0=\pi/9$ and $b=6$ in Eq. (\ref{PD}) gives $P_{\rm D}(6)\approx
2$, which is insufficient for such dislocations to be bound in
pairs \cite{NHB}. The application of the Debye-H\"{u}ckel
approximation to the two-dimensional Coulomb gas formed by
dislocations shows that $c_{\rm D}$, the concentration of free
dislocations, should be exponentially small in $1/T$.
% depend on temperature as $c_{\rm D}\propto Y^{4/(4-P_{\rm D})}$,
% where $Y=\exp(-E_{\rm D}/T)$ is the fugacity and $E_{\rm D}\sim J$
% the core energy of a dislocation.
The value of $c_{\rm D}$ determines a temperature dependent
correlation radius $r_c(T)\propto c^{-1/2}_{\rm D}$, beyond which
$K$ is renormalized to zero and even the algebraic correlations of
vortex pattern are destroyed.

The presence of free dislocations leads to the screening of the
entropic part of the logartihmic interaction of half-vortices at
the scales which are large in comparison with $r_c(T)$.
Nonetheless, at low temperatures the system will be characterized
by a finite value of $\Gamma(T)$, since all half-vortices will be
bound in pairs by their direct interaction. With increasing
temperature a phase transition will occur related to appearance of
free half-vortices and vanishing of $\Gamma(T)$. It will be
completely analogous to the
Berezinskii-Kosterlitz-Thouless phase transition %\cite{Ber,KT,Kost}
in the conventional $XY$ model (without frustration), the main
difference being that half-vortex pairs dissociate when
$T=(\pi/8)\Gamma(T)$ \cite{K86c}, whereas the pairs of ordinary
vortices dissociate when $T=(\pi/2)\Gamma(T)$ \cite{NK}.

Thus, in the present work we have shown that in the frustrated
$XY$ model with a dice lattice and $f=1/3$ the vortex pattern is
disordered at any temperature (becoming quasi-ordered only at
$T=0$). Nonetheless, at low temperatures the helicity modulus is
finite and jumps to zero only at $T=T_{\rm HV}\sim
(\pi/8)\Gamma_0\approx 0.28\,J$, where the pairs of half-vortices
dissociate. This estimate is not far from the value of the
transition temperature, $T_c\approx 0.2\,J$, obtained in numerical
simulations of Ref. \onlinecite{CF}.

A possibility of vortex pattern ordering appears only when one
goes beyond the limits of the $XY$ model and takes into account
some additional mechanism of the removal of an accidental
degeneracy. In the case of a proximity coupled array \cite{TTM}
the main role will belong to the energy related to the magnetic
fields of currents in the array, which is minimized in the striped
state \cite{footnote2}. In terms of the SOS model this leads to a
negative step energy, $E_{\rm st}<0$. In the limit of weak
screening, when the corrections to currents from their proper
magnetic fields can be neglected, $E_{\rm st}=-\mu J^2/E_{\Phi}$,
where $E_{\Phi}=\Phi_0^2/4\pi^2 a$ is the characteristic energy,
$a$ is the lattice constant (of a dice lattice) and the numerical
coefficient $\mu$ for both types of steps shown in Fig. 2(d) can
be written as
$$
\mu\approx\sin^2(\pi/3)\cdot
[\lambda_2-\lambda_4-2(\lambda_5-\lambda_6)+\ldots]\approx 0.25\;,
$$
$\lambda_i\equiv-L_i /a>0$ being the dimensionless values of
mutual inductances, $L_i$,  between dice lattice plaquettes
\cite{nwd}. For $a\approx 8\,\mu{\rm m}$ \cite{TTM}
$E_{\Phi}\approx 10^4\,{\rm K}$.
% Another mechanism for the
% appearance of a finite $E_{\rm st}$ consists in taking into
% account the energy of quantum fluctuations.

With decrease of $T$ the ratio $|E_{\rm st}|/T$ is increased,
which for  $|E_{\rm st}|/T\ll 1$ will manifest itself only in the
decrease of $K$ in Eq. (\ref{Hh}). With further decrease of $T$ a
phase transition can be expected to occur to a phase with a
non-zero slope. In terms of vortices this phase will be
characterized by a true long-range order manifesting itself in
deviation of occupation probabilities for plaquettes with
different orientations from $1/3$. This phase transition can be
expected to happen when $|E_{\rm st}|\sim T$, {\em i.e.},
$T/J\sim(\mu
T/E_{\Phi})^{1/2}\sim 10^{-2}$, % that is well below $T_{\rm HV}$,
and, according to the results of Ref. \onlinecite{NKL}, has to be
of the first order. However, at $T/J\lesssim 0.05$ the relaxation
of vortex pattern is likely to be dynamically quenched [at
$T/J=0.05$ one has $\exp(-E_{\rm b}/T)\sim 10^{-14}$], which may
prevent the observation of such an ordering in experiments or
simulations.

\vspace*{2mm}

The author is grateful to P. Martinoli for useful comments. This
work has been supported in part by the Program "Quantum
Macrophysics" of the Russian Academy of Sciences, by the Program
"Scientific Schools of the Russian Federation" (grant No.
1715.2003.2), and also by the Swiss National Science Foundation.


\begin{thebibliography}{99}
\bibitem{footnote1} In the wire network problem with the same geometry this
state is selected already when one compares the harmonic
contributions to the Ginzburg-Landau functional.
\bibitem{footnote2} This is likely to be the main reason for the
observation of this state in the decoration experiments on
superconducting wire networks \cite{Serr}.
\end{thebibliography}
\end{document}